\documentstyle[aps,twocolumn]{revtex}
\begin{document}
\title{Robustness of the van Hove scenario for high-$T_c$ superconductors}
\author{V.Yu.Irkhin$^{*}$, A.A.Katanin, and M.I.Katsnelson}
\address{Institute of Metal Physics, 620219, Ekaterinburg, Russia}
\maketitle

\begin{abstract}
The pinning of the Fermi level to the van Hove singularity and the formation
of flat bands in the two-dimensional $t$-$t^{\prime }$ Hubbard model is
investigated by the renormalization group technique. The ``van Hove''
scenario of non-Fermi-liquid behavior for high-$T_c$ compounds can take
place in a broad enough range of the hole concentrations. The results are in
qualitative agreement with the recent ARPES data on La$_2$CuO$_4$.
\end{abstract}


The nature of normal state and the mechanisms of superconductivity in
copper-oxide high-$T_c$ compounds are still a subject of hot discussions 
\cite{Anderson,Anderson1,Scalapino,Pines,SO5,Dagotto}. A number of
experimental data on transport properties and ARPES give an evidence of a
non-Fermi-liquid (NFL) character of the normal state in underdoped regime
and, as a consequence, of a strict confinement of current carriers in the CuO%
$_2$ planes. The latter results in an anomalously weak sensitivity of
normal-phase and superconducting properties to the impurities outside the
planes, hopping character of the transport along $c$-axis etc. Anderson \cite
{Anderson} has put forward the idea that a 2D system can demonstrate a NFL
behavior at arbitrarily small interelectron repulsion $U$ owing to a finite
phase shift at the Fermi energy. In this scenario the confinement of current
carriers is explained by ``quantum protection'' \cite{Anderson1} owing to
charge-spin separation and incoherent (non-quasiparticle) character of
electron (or hole) motion. However, theoretical understanding of NFL state
in quasi-2D electron systems turned out to be very difficult. Modern
field-theoretical investigations \cite{Nest} show that such a state,
probably, does not occur in general 2D and quasi-2D cases.

At the same time, peculiarities of electron spectrum can lead to NFL
behavior. In the presence of van Hove singularities (VHS) near the Fermi
level the marginal Fermi-liquid \cite{MF} or NFL behavior \cite
{Dzyaloshinskii} can be naturally derived. Already in the leading order of
the perturbation theory in $U$ the marginal dependences of the electron
scattering rate and specific heat take place, ${\rm Im}\Sigma ({\bf k}%
_F,\varepsilon )\propto |\varepsilon |\ln |1/\varepsilon |\ $at $%
|\varepsilon |\gg |\mu |$ and $\delta C\propto T\ln ^3(t/\max (|\mu |,T)),$
and the resistivity demonstrates the behavior $\rho \propto T\ln ^2(t/\max
(|\mu |,T))$ \cite{OurVH1} ($T$ is the temperature, $\varepsilon $ is the
energy, $\mu $ is the chemical potential calculated from VHS, and $t$ is the
hopping integral). Closeness of the VHS to the Fermi level results in a
1D-like behavior of the perturbation expansion \cite{Rice0}, but does not
lead to the Luttinger-liquid fixed point since the system turns out to be
unstable with respect to formation of magnetic or superconducting ground
state. It has been shown \cite{Rice0,Rice,OurVH} in the framework of 2D $t$-$%
t^{\prime }$ Hubbard model that the closeness of the Fermi surface to VHS
leads to the instabilities of the Fermi-liquid state in the finite range of
the concentrations near the van Hove filling. The type of the instability
(magnetic or superconducting) depends on the value of $t^{\prime }/t$.

This ``van Hove scenario'' (for a review of early considerations, see, e.g.,
Ref.\cite{VHS}) seems to be very attractive since various data demonstrate
the closeness of VHS to the Fermi level in high-$T_c$ compounds at optimal
doping or pressure. Recent ARPES data demonstrate that the Fermi level of La$%
_{2-x}$Sr$_x$CuO$_4$ is close to VHS in the range $x=0.20\div 0.30$ \cite
{Ino}, which corresponds to overdoped regime. One of the two bands, that
occur due to bilayer splitting (antibonding band) in Bi2212 system, is also
close to VHS \cite{Feng}. A significant and still unsolved question is why
the Fermi level should be close to VHS in a broad doping range. Although
some considerations were performed \cite{kastr,Gonzalez}, they did not take
into account the renormalization of electron spectrum, e.g., the formation
of the flat part of the spectrum, which should change substantially the
results of these approaches.

The flattening of the spectrum was first studied theoretically for a single
hole in an antiferromagnetic background \cite{Hole}. The flat region leads
to some peculiarities of electron properties near the metal-insulator
transition \cite{MIT}, in particular to the pinning of the chemical
potential. While above-mentioned approaches describe correctly the situation
at small hole doping (the doped Mott insulator regime), these approaches
meet with difficulties near the optimal doping where the concentration of
holes is not small. In this Letter we argue that besides antiferromagnetic
fluctuations there is another factor which results in occurrence of the flat
part of electron spectrum, namely, the critical fluctuations that occur near
the van Hove band filling. We also reconsider the issue of the pinning of
Fermi surface to van Hove singularities by proposing a new scheme of
renormalizations of the energy- and momentum-dependent electron Green's
function. We show that the pinning of the Fermi surface to VHS\ is an
universal feature of 2D systems which is connected with the formation of a
flat region of electron spectrum near the ($\pi ,0$) point. Both the
phenomena, the pinning and flattening, will be described within the same
renormalization group (RG) approach.

We start from the $t$-$t^{\prime }$ Hubbard model on the square lattice, 
\begin{equation}
H=\sum_{{\bf k}}\varepsilon _{{\bf k}}c_{{\bf k}\sigma }^{\dagger }c_{{\bf k}%
\sigma }+U\sum_in_{i\uparrow }n_{i\downarrow }  \label{H}
\end{equation}
with the bare electron spectrum 
\begin{equation}
\varepsilon _{{\bf k}}=-2t(\cos k_x+\cos k_y)-4t^{\prime }(\cos k_x\cos
k_y+1)-\mu  \label{ek}
\end{equation}
Hereafter we assume $t>0,$ $t^{\prime }<0$ (which is the case for hole-doped
systems), $0\leq |t^{\prime }|/t<1/2$. For arbitrary $t^{\prime }/t$ the
spectrum (\ref{ek}) contains VH singularities connected with the points $%
A=(\pi ,0),$ $B=(0,\pi ).$ The chemical potential $\mu $ measures the
distance between VHS and the Fermi energy, so that at $\mu =0$ VHS lie at
the Fermi level. We suppose that $t^{\prime }/t$ is not too small, and
therefore the Fermi surface is not nested. It was argued \cite
{Dzyaloshinskii,Rice,OurVH} that provided that the band filling is close to
van Hove one, the vicinity of the van Hove points gives a dominant (most
singular) contribution to electronic and magnetic properties. Thus we can
use the expansion of the spectrum near $A(B)$ van Hove points 
\begin{mathletters}
\begin{eqnarray}
\varepsilon _{{\bf k}}^A &=&-2t(\sin ^2\varphi \overline{k}_x^2-\cos
^2\varphi k_y^2)-\mu  \label{eka} \\
\varepsilon _{{\bf k}}^B &=&2t(\cos ^2\varphi k_x^2-\sin ^2\varphi \overline{%
k}_y^2)-\mu  \label{ekb}
\end{eqnarray}
where $\overline{k}_x=\pi -k_x,$ $\overline{k}_y=\pi -k_y,\varphi $ is the
half of the angle between asymptotes at VHS, $2\varphi =\cos
^{-1}(-2t^{\prime }/t).$ The chemical potential $\mu $ is obtained from the
electron concentration $n=\int\nolimits_{-\Lambda t}^0\rho (\varepsilon
)d\varepsilon $ with $\rho (\varepsilon )$ being the renormalized density of
states$,$ $\Lambda $ the ultraviolet cutoff. For the spectrum (\ref{ek}) we
have the bare density of states 
\end{mathletters}
\begin{equation}
\rho _0(\varepsilon )=\left( \pi ^2\sqrt{t^2-(2t^{\prime })^2}\right)
^{-1}\ln \frac{\Lambda t}{|\varepsilon +\mu |}  \label{r0}
\end{equation}
which diverges logarithmically at $\varepsilon =-\mu $. We will show that
the renormalized density of states contains a much stronger divergence,
which results in the pinning of the Fermi surface to VHS. To this end we
calculate the electron self-energy $\Sigma ({\bf k},\varepsilon )$ for ${\bf %
k}$ near the VH point ${\bf k}_{VH}=(\pi ,0).$ In the second order in $U$
this contains three contributions that arise from intermediate quasimomenta
which are close to the same or another VH point (cf. Refs.\cite
{Dzyaloshinskii,OurVH1}), $\Sigma ^{(2)}({\bf k},\varepsilon
)=\sum_{i=1}^3\Sigma _i({\bf k},\varepsilon )$ with 
\begin{eqnarray}
\text{Re}\Sigma _1({\bf k},\varepsilon ) &=&-(2g_0/\sin 2\varphi
)^2[A_1\varepsilon +B_1^x\widetilde{k}_x^2  \nonumber \\
&&\ \ \ \ \ \ \ \ \ \ \ \ \ \ \ \ \ \ \ \ \ \ \ +B_1^y\widetilde{k}_y^2]\ln
^2(\Lambda /\upsilon )  \nonumber \\
\text{Re}\Sigma _{2,3}({\bf k},\varepsilon ) &=&-2(g_0/\sin 2\varphi
)^2[A_{2,3}\varepsilon +B_{2,3}^x\widetilde{k}_x^2  \nonumber \\
&&\ \ \ \ +2C_{2,3}\widetilde{k}_x\widetilde{k}_y+B_{2,3}^y\widetilde{k}%
_y^2]\ln (\Lambda /\upsilon )  \label{S2}
\end{eqnarray}
where $\widetilde{k}_x=\overline{k}_x\sin \varphi ,$ $\widetilde{k}%
_y=k_y\cos \varphi ,$ $g_0=U/(4\pi ^2t)$ is the dimensionless coupling
constant, $\upsilon =\max (\overline{k}_x^2,k_y^2,|\varepsilon |/2t,|\mu
|/2t),\;A_1=\ln 2,$ $B_1^x=-B_1^y=1/2-\ln 2.$ The analytical expressions for
the coefficients $A_i,B_i$, and $C_i$ with $i=2,3,$ which are some regular
functions of $t^{\prime }/t,$ will be published elsewere. We perform a
summation of logarithmically divergent terms which depend on $\lambda
=(1/2)\ln (\Lambda /\upsilon )$ within the RG approach. To separate effects
of the electron spectrum renormalization and multiplicative renormalization
of the Green's function, we perform the renormalization procedure in two
steps. At the first step we introduce the{\bf \ }${\bf k}$-dependent mass
renormalization factors $Z_{mi}^a(\upsilon )$ ($a=x,y$), and at the second
one all the other divergences are absorbed into the energy-dependent
quasiparticle residue $Z(\varepsilon ),$%
\begin{equation}
G({\bf k},\varepsilon )=\frac{Z(\varepsilon )}{\varepsilon
+2tZ_{xm}^{-1}(\upsilon )\widetilde{k}_x^2-2tZ_{ym}^{-1}(\upsilon )%
\widetilde{k}_y^2+\widetilde{\mu }+i\Gamma (\varepsilon )}  \label{gf}
\end{equation}
where $\widetilde{\mu }=\mu +{\rm Re}\Sigma ({\bf k}_{VH},-\mu )$ is the
renormalized chemical potential, the damping $\Gamma (\varepsilon
)=Z(\varepsilon ){\rm Im}\Sigma ({\bf k}_{VH},\varepsilon )$ is determined
by analytical continuation. The coefficients $Z_{am}(\upsilon
)=Z_{m1}^a(\upsilon )Z_{m2}^a(\upsilon )Z_{m3}^a(\upsilon )$ satisfy the RG
equations 
\begin{eqnarray}
\frac{d\ln Z_{m1}^a(\lambda )}{d\lambda } &=&(B_1^a-A_1)\lambda \gamma
_4^2/\sin ^22\varphi  \nonumber \\
\frac{d\ln Z_{m2}^a(\lambda )}{d\lambda } &=&(B_2^a-A_2)(\gamma _1^2+\gamma
_2^2-\gamma _1\gamma _2)/\sin ^22\varphi  \nonumber \\
\frac{d\ln Z_{m3}^a(\lambda )}{d\lambda } &=&(B_3^a-A_3)\gamma _3^2/\sin
^22\varphi  \label{ZZ}
\end{eqnarray}
Here $\gamma _i$ are four-electron vertices determined from 
\begin{eqnarray}
d\gamma _1/d\lambda &=&4d_1(\lambda )\gamma _1(\gamma _2-\gamma
_1)+4d_2\gamma _1\gamma _4-4\,d_3\gamma _1\gamma _2  \nonumber \\
d\gamma _2/d\lambda &=&2d_1(\lambda )(\gamma _2^2+\gamma _3^2)+4d_2(\gamma
_1-\gamma _2)\gamma _4-2d_3(\gamma _1^2+\gamma _2^2)  \nonumber \\
d\gamma _3/d\lambda &=&-4d_0(\lambda )\gamma _3\gamma _4+4d_1(\lambda
)\gamma _3(2\gamma _2-\gamma _1)  \nonumber \\
d\gamma _4/d\lambda &=&-2d_0(\lambda )(\gamma _3^2+\gamma _4^2)+2d_2(\gamma
_1^2+2\gamma _1\gamma _2-2\gamma _2^2+\gamma _4^2)  \label{TwoPatch}
\end{eqnarray}
where 
\begin{eqnarray}
d_0(\lambda ) &=&\lambda (1-R^2)^{-1/2};  \nonumber \\
d_1(\lambda ) &=&\min (\lambda ,\ln [(1+\sqrt{1-R^2})/R])  \nonumber \\
d_2 &=&(1-R^2)^{-1/2};\;d_3=\tan ^{-1}(R/\sqrt{1-R^2})/R
\end{eqnarray}
with $R=-2t^{\prime }/t.$ Equations for $Z(\upsilon )=Z_1(\upsilon
)Z_2(\upsilon )Z_3(\upsilon )$ are obtained by the replacement $%
B_i^a-A_i\rightarrow A_i$ in (\ref{ZZ}). Although the renormalization-group
equations (\ref{ZZ}) and (\ref{TwoPatch}) are only approximate because of
the presence of $\ln ^2(\Lambda /\upsilon )$ terms in the series expansions,
it was argued in Ref. \cite{OurVH} that they reproduce correctly the
solutions of the parquet equations which take into account complete momentum
dependence of the vertices. Numerical solution of Eqs. (\ref{ZZ})
demonstrates that the vertices $\gamma _i$ increase with decreasing $%
\upsilon $ and diverge at the critical value $\upsilon =\upsilon
_c(t^{\prime }/t)$ \cite{OurVH}$.$ We have $Z_{am}(\upsilon )\rightarrow
\infty $ at $\upsilon \rightarrow \upsilon _c$ while $Z_a(\upsilon
)\rightarrow 0.$ This divergence signals a transition into magnetically
ordered or superconducting state at $|\mu |<\mu _c=2t\upsilon _c.$ The
equations (\ref{ZZ}), (\ref{TwoPatch})\ are valid in weak- and
intermediate-coupling regimes where $\gamma _i<1$, i.e. at $|\mu |$ not too
close to $\mu _c.$

The electron spectrum is determined by the pole of the Green's function (\ref
{gf}). Increasing the factors $Z_{am}(\upsilon )$ with decreasing $\upsilon $
leads to $|\varepsilon +\mu |<2t\min (\overline{k}_x^2,k_y^2)$ at the pole,
so that $\upsilon =\max (\overline{k}_x^2,k_y^2,|\widetilde{\mu }|/2t)$ to
logarithmic accuracy. This demonstrates that the separation of momentum- and
energy-renormalization effects turns out to be self-consistent. We obtain
therefore for the renormalized spectrum $E_{{\bf k}}$%
\begin{equation}
E_{{\bf k}}=-\widetilde{\mu }+2t\left\{ 
\begin{array}{cc}
Z_{ym}^{-1}(k_y^2)\widetilde{k}_y^2-Z_{xm}^{-1}(k_x^2)\widetilde{k}_x^2, & 
k_{x,y}^2>|\widetilde{\mu }|/(2t) \\ 
Z_{ym}^{-1}(|\widetilde{\mu }|/2t)\widetilde{k}_y^2-Z_{xm}^{-1}(|\widetilde{%
\mu }|/2t)\widetilde{k}_x^2 & k_{x,y}^2<|\widetilde{\mu }|/(2t)
\end{array}
\right.  \label{Ek}
\end{equation}
For $|\widetilde{\mu }|>\mu _c$ and $k\gg k_c=\upsilon _c^{1/2}$ we have $%
Z_{xm,ym}^{-1}(k^2)\simeq 1,$ and the dispersion coincides with the bare
one, while at $\widetilde{k}_x,\widetilde{k}_y<k_c$ the spectrum becomes
more flat due to renormalization by the factors $Z_{xm,ym}^{-1}(|\widetilde{%
\mu }|/2t).$ Although the regime $|\widetilde{\mu }|\leq \mu _c,$ $k<k_c$ is
outside of the region of the validity of RG equations (\ref{ZZ}) and (\ref
{TwoPatch}), we have formally $E_{{\bf k}}=-\widetilde{\mu }$ for $k<k_c,$
i.e. the spectrum is flat in this region. This is similar to the result of
Dzyaloshinskii \cite{Dzyaloshinskii}. However, unlike that paper, we have
taken into account all the channels of electron scattering, which gives a
flat part of the spectrum already in one-loop approximation for the
vertices. The flat part formation was considered earlier as one of formally
possible instability channels of Landau Fermi-liquid (``fermionic
condensation'' \cite{Khodel,nozieres}). Here we demonstrate that similar
phenomena can take place in two-dimensional systems near the van Hove
filling. The result of numerical calculation of the dispersion law $E_{{\bf k%
}}$ according to Eqs. (\ref{ZZ}) and (\ref{Ek}) at different values of $%
\widetilde{\mu }$ is shown in Fig.1. One can see that even at $|\widetilde{%
\mu }|>\mu _c$ the spectrum has a wide almost flat region$.$ The picture
shown in Fig.1 is in a good qualitative agreement with the ARPES data for La$%
_{2-x}$Sr$_x$CuO$_4$ \cite{Ino}. These data give a possibility to estimate $%
k_c\simeq 0.6a^{-1}$ ($a$ is the lattice constant) for this system;
according to (\ref{ek}), this $k_c$ value corresponds to $\widetilde{\mu }%
_c/(2t)=\upsilon _c\simeq 0.1.$ We stress once more the difference of the
present approach with the approaches \cite{AFCorr} which yield the
flattening of the spectrum owing to strong antifferomagnetic correlations
and therefore are reliable only at low hole concentrations.

The flat part of electron spectrum leads to drastic changes in the density
of states and dependence $n(\widetilde{\mu })$. The contribution of the flat
part of the spectrum to the density of states can be written as 
\begin{eqnarray}
\delta \rho (\varepsilon ) &=&-\frac{k_c^2}{\pi ^2}\frac{\text{Im}\overline{%
\Sigma }({\bf k}_{VH},\varepsilon )}{[\varepsilon +\widetilde{\mu }-\text{Re}%
\overline{\Sigma }({\bf k}_{VH},\varepsilon )]^2+[\text{Im}\overline{\Sigma }%
({\bf k}_{VH},\varepsilon )]^2}  \nonumber \\
\ &=&\frac{k_c^2}\pi A({\bf k}_{VH},\varepsilon )
\end{eqnarray}
where $\overline{\Sigma }({\bf k},\varepsilon )=\Sigma ({\bf k},\varepsilon
)-{\rm Re}\Sigma ({\bf k},-\mu )$ and $A({\bf k}_{VH},\varepsilon )$ is the
quasiparticle spectral weight at the van Hove momentum.

Consider first the results obtained within the second-order expression for $%
\Sigma ({\bf k}_{VH},\varepsilon ),$ Eq. (\ref{S2}). Note that at this stage
the effects connected with the renormalization of electron dispersion are
already taken into account by performing summation of ${\bf k}$-dependent
logarithmically divergent terms to all orders of perturbation theory. We
have at $\;|\widetilde{\mu }|\ll |\varepsilon |\ll t$ to leading logarithmic
accuracy for $\overline{\Sigma }({\bf k}_{VH},\varepsilon )$ the result 
\begin{equation}
\delta \rho (\varepsilon )\simeq \frac{k_c^2}{\pi |\varepsilon |}\frac{C\ln
(\Lambda t/|\varepsilon |)}{[1+C\ln ^2(\Lambda t/|\varepsilon |)]^2},\;C=%
\frac{g_0^2\ln 2}{\sin ^22\varphi }.
\end{equation}
Therefore we obtain 
\begin{eqnarray}
n(\widetilde{\mu }) &=&n_{VH}-\frac{\widetilde{\mu }}{\pi ^2\sqrt{%
t^2-(2t^{\prime })^2}}\ln \frac{\Lambda t}{|\widetilde{\mu }|}  \label{nm} \\
&&\ \ \ \ \ \ \ \ \ -\frac{k_c^2}{2\pi }\frac{{\rm sign}\text{ }\widetilde{%
\mu }}{1+C\ln ^2(\Lambda t/|\widetilde{\mu }|)},  \nonumber
\end{eqnarray}
where $n_{VH}$ is the van Hove filling. The second term in the right-hand
side of Eq. (\ref{nm}) comes from momenta $k>k_c$ outside the flat part of
the spectrum and can be calculated by using the bare density of states (\ref
{r0}) since the renormalization of the spectrum at these momenta is not too
important. For small enough $|\widetilde{\mu }|$ we can neglect this term in
comparison with the third term to obtain 
\begin{equation}
n(\widetilde{\mu })=n_{VH}-\frac{k_c^2\sin ^22\varphi }{2\pi g_0^2\ln 2}%
\frac{{\rm sign}\text{ }\widetilde{\mu }}{\ln ^2(\Lambda t/|\widetilde{\mu }%
|)},
\end{equation}
(the unity is small in comparison with squared logarithm in the
denominator). The chemical potential measured from the VHS energy is given
by a non-analytical function, 
\begin{equation}
\widetilde{\mu }(n)=\Lambda t\exp (-{\rm const}/|n-n_{VH}|^{1/2}),
\label{nm1}
\end{equation}
and is therefore practically zero in a rather wide range of electron
concentrations $n$ near $n_{VH}.$ Generally speaking, in the absence of the
quadratic terms in the electronic spectrum, quartic terms may become
important. However, the presence of such terms give only subleading
corrections to the above results and does not change them qualitatively.

The results of numerical calculations of $n(\widetilde{\mu })$ with Eq. (\ref
{nm}) and the renormalized dependence $n^{\prime }(\widetilde{\mu })$ with
account of renormalization of $\Sigma ({\bf k}_{VH},\varepsilon )$ according
to Eq. (\ref{ZZ}) are shown in Fig.2 for $U=4t$ and $k_c\simeq 0.6$. The
dependence $n_0(\widetilde{\mu })$ obtained by integrating the bare density
of states (\ref{r0}) is also shown for comparison. The result for $n^{\prime
}(\widetilde{\mu })$ is shown at $\widetilde{\mu }<-\mu _c$ only. One can
see that the dependence $n(\widetilde{\mu })$ yields the pinning of the
Fermi surface in the concentration range about $4\%$ above and below VH
filling (in fact, in the present approach the picture is symmetric, $n(%
\widetilde{\mu })-n_{VH}=n_{VH}-n(-\widetilde{\mu })$). Such a behavior is
in a qualitative agreement with the ARPES results of Ref. \cite{Ino}. On the
other hand, this is in contrast with the behavior of chemical potential
measured relative to its position in the insulating phase, Ref.\cite{Ino1}.

It is important that the condition $|n^{\prime }(\widetilde{\mu }%
)-n_{VH}|>|n(\widetilde{\mu })-n_{VH}|$ holds.{\it \ }This means that after
account of renormalizations the pinning effect becomes even stronger than
that given by Eqs.(\ref{nm}) and (\ref{nm1}). Being extrapolated to the
region $|\widetilde{\mu }/t|<\upsilon _c,$ the dependence $n^{\prime }(%
\widetilde{\mu })$ should give even larger values of critical concentrations
(about $6\%$). One can expect that for larger values of $U/t$ the critical
concentrations will be larger.

To conclude, we have developed a RG approach which gives a possibility to
describe the flattening of the electron spectrum and does not suppose the
presence of strong antiferromagnetic fluctuations. As a result of the
peculiar structure of the spectrum, the pinning of the Fermi level to van
Hove singularities in 2D systems occurs, the chemical potential being
practically constant in a range of dopings near VH filling. In the pinning
region the electron density of states is determined by the quasiparticle
damping and the system demonstrates essentially non-Fermi-liquid behavior.
Further experimental investigations on LaSrCuO and Bi2212 systems would be
of interest to verify the pinning picture proposed.

We are grateful to Arno Kampf for helpful discussions. The research
described was supported in part by Grant No.00-15-96544 from the Russian
Basic Research Foundation and by Russian Science Support Foundation.

{\sc Figure captions}

Fig.1. Quasiparticle dispersion for $t^{\prime }/t=-0.3$ and $U=4t$ from RG
approach. The values of the chemical potential are $\widetilde{\mu }%
=0,-0.2t,-0.4t$ (from top to bottom).

Fig.2. The dependence $n(\widetilde{\mu })-n_{VH}$ for the same parameter
values as in Fig.1, $\widetilde{\mu }$ being referred to VHS energy. The
adopted value of $k_c$ is $0.6.$ The dashed line corresponds to bare
electron spectrum, and the solid line to the result (\ref{nm}). The
dot-dashed line is the dependence with account of renormalizations of $%
\Sigma ({\bf k}_{VH},\varepsilon ).$

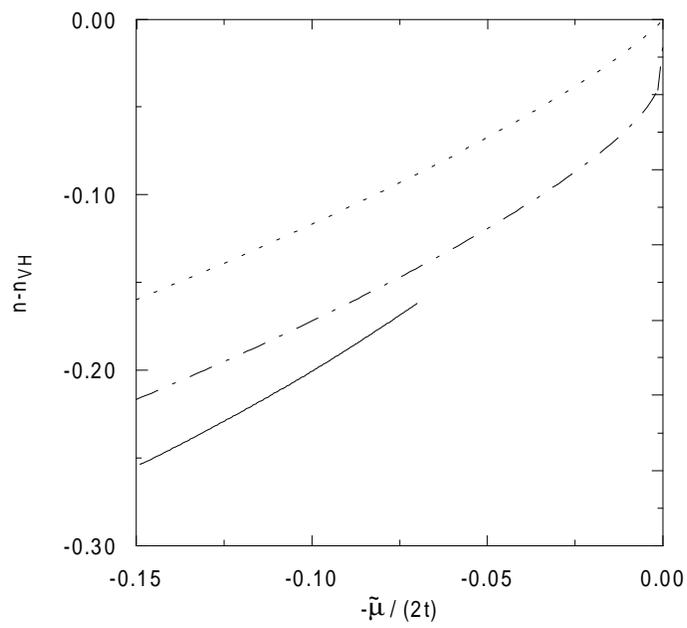
\end{document}